# Predicting Phase Transitions in PbTiO$_3$ using Zentropy through Quasi-Harmonic Phonon Calculations

Nigel Lee En Hew[1,*], Shun-Li Shang[1], and Zi-Kui Liu[1]

[1]Department of Materials Science and Engineering, The Pennsylvania State University, University Park, Pennsylvania 16802, USA

[*]Corresponding author: njh5724@psu.edu

ORCID: 0000-0003-1374-4589 (Nigel Lee En Hew); 0000-0002-6524-8897 (Shun-Li Shang); 0000-0003-3346-3696 (Zi-Kui Liu)

**Abstract**

According to X-ray diffraction (XRD) measurements, PbTiO$_3$ undergoes a phase transition from a tetragonal ferroelectric phase to a cubic paraelectric phase at 763 K. However, X-ray absorption fine-structure (XAFS) measurements indicate that PbTiO$_3$ is locally tetragonal even after the phase transition. The difference in these results is because XAFS measurements can probe local features of a structure, while XRD averages over such local features. For both measurements to be consistent, PbTiO$_3$ is macroscopically cubic but locally tetragonal after the phase transition. Despite this, most models, such as the Laundau-Ginsburg-Devonshire theory and effective Hamiltonians, are still unable to explain this phenomenon. Moreover, these methods involve model parameters fitted to experimental or theoretical data and do not consider other tetragonal configurations, such as domain walls, to predict the phase transition. Our previous study used our novel zentropy approach to predict the phase transition by considering the tetragonal ferroelectric ground state configuration and the tetragonal 90° and 180° domain wall configurations with their total energies at 0 K. In the present work, the Helmholtz energies of the three configurations are obtained from density functional theory calculations through energy-volume curves and phonon calculations. The predicted phase transition temperature using the meta-GGA r$^2$SCAN and revised multiplicities of configurations is 716 K, showing good agreement with the experimental value of 763 K.

# 1. Introduction

Ferroelectric (FE) materials possess a spontaneous polarization without an applied electric field, which can be reversed by an externally applied electric field [1]. Lead titanate (PbTiO$_3$) is a well-known ferroelectric material with a perovskite crystal structure [1] with applications in nonvolatile random-access memories, sensors, actuators, and transducers [2–5]. At T < 763 K, PbTiO$_3$ adopts a tetragonal structure with a c/a ratio = 1.064 that exhibits ferroelectricity (**Figure 1a**). At T > T$_c$ = 763 K without external stress and an applied electric field, PbTiO$_3$ undergoes a phase transition to a cubic structure that is paraelectric (PE) (**Figure 1b**). These results are based on X-ray diffraction (XRD) measurements that provide average structural information of a material [6]. However, x-ray absorption fine-structure (XAFS) measurements [7–9] that can probe local structural information at the atomic scale with a very short effecting measuring time of ~ 10$^{-16}$ s found that a c/a ratio exists throughout the temperature range tested, including at T > T$_c$. These observations indicate that PbTiO$_3$ is still locally tetragonal after the phase transition. Ab initio molecular dynamics simulations (AIMD) conducted by Fang et al. [10] found similar results. Thus, for the results of both XRD and XAFS measurements to be consistent, we hypothesize that PbTiO$_3$ is macroscopically cubic but still tetragonal at the microscopic level at T > T$_c$.

The FE and PE behaviors of PbTiO$_3$ are commonly described by the phenomenological Landau-Ginsburg-Devonshire theory (LGDT) [11], which involves fitting model parameters to experimental observations [12]. LGDT postulates that the macroscopic polarization changes smoothly from a finite value in the FE phase to zero in the PE phase, which agrees well with macroscopic experimental measurements [11,12]. However, this method cannot explain the microscopic features observed in XAFS experiments [7–9]. On the other hand, the modern theory of polarization (MTP) [13,14] focuses on the differences in polarization between two different states, which is what polarization-reversal experiments measure. One critical result of MTP compared to LGDT is that the microscopic polarization does not necessarily vanish in the PE phase; that is, the Ti-O bond lengths can remain unequal [13–15]. Other approaches include effective Hamiltonians and various potential approaches, which have been developed to predict the FE-PE transitions at finite temperatures. The model parameters are first fitted to density functional theory (DFT) calculations, followed by Monte Carlo or molecular dynamic simulations [16–20]. The predictions from those simulations agree well with experiments regarding the phase stability sequence, transition temperatures, latent heat, and spontaneous polarizations. They also provide valuable insights into the order-disorder versus the displacive character of the phase transition [8,16,17,21]. While there are advantages to these methods, they all involve model parameters fitted to either experimental or theoretical data, which reduces the predictability of these approaches.

In this study, we apply our zentropy approach to study the FE-PE phase transition in $PbTiO_3$. Contrary to the abovementioned methods, this novel approach does not require any fitting parameters and takes inputs only from DFT calculations, similar to ab initio molecular dynamics (AIMD) simulations. However, Zentropy differs from AIMD in that you can include as many relevant configurations as practically possible. In contrast, AIMD may not be able to sample some relevant configurations due to the simulation time limitations and the shape of the supercell. Moreover, Zentropy allows us to compute properties like the configurational entropy due to the statistical mixing of configurations, which is not straightforward in AIMD.

Zentropy has been successfully used to predict magnetic transitions in Ce [22–24], bcc Fe [25], fcc Ni [26], $BaFe_2As_2$ [27], $Fe_3Pt$ [24,28,29], and $YNiO_3$ [30]. More recently, Liu et al. predicted the $T_c$ of $PbTiO_3$ using the domain wall energies of the 90° domain wall (90DW) and the 180° domain wall (180DW) energies at 0 K from the literature, which was computed using DFT with LDA as the exchange-correlation functional [31]. The present study extends the previous work as we have calculated the free energy of each configuration and used it to obtain other relevant properties.

## 2. Overview of Zentropy

The full details for zentropy are mentioned in the previous citations for magnetic materials and $PbTiO_3$ and other recent publications [28,32–34]. Here, we reiterate the relevant theory and equations. We start with the definition of the partition function $Z^k$ of a particular configuration, k

$$Z^k = e^{\frac{-F^k}{k_B T}} \quad \text{Eq. 1}$$

where $F^k$ is the free energy of a configuration k, $k_B$ is the Boltzmann constant, and T is the temperature. Then, the partition function for a system, Z, that consists of m configurations is

$$Z = \sum_{k=1}^{m} e^{\frac{-F^k}{k_B T}} = \sum_{k=1}^{m} Z^k = e^{\frac{-F}{k_B T}} \quad \text{Eq. 2}$$

The last equality in **Eq. 2** comes from F = -k$_B$T ln Z, where F is the free energy of the system. Next, the probability of a configuration, p$^k$, is given by

$$p^k = \frac{e^{\frac{-F^k}{k_B T}}}{Z} = \frac{Z^k}{Z} \quad \text{Eq. 3}$$

Note that **Eq. 1** to **Eq. 3** only differs from standard statistical mechanics by substituting F$^k$ with E$^k$, the internal energy of each configuration. The use of F$^k$ is specific to Zentropy, where we assume each configuration has its own entropy S$^k$, given by F$^k$ = E$^k$ – TS$^k$. If you assume that S$^k$ is equal to zero, **Eq. 1** to **Eq. 3** revert to standard statistical mechanics.

The free energy of the system, F, can then be derived as follows, taking advantage of the fact that the sum of p$^k$ for all the configurations is always equal to 1

$$F = -k_B T \ln Z$$

$$F = -k_B T \sum_{k=1}^{m} p^k \ln Z$$

$$F = -k_B T \sum_{k=1}^{m} p^k \ln Z + k_B T \sum_{k=1}^{m} p^k \ln Z^k - k_B T \sum_{k=1}^{m} p^k \ln Z^k$$

$$F = \sum_{k=1}^{m} p^k F^k + k_B T \sum_{k=1}^{m} p^k \ln p^k \quad \text{Eq. 4}$$

Using the fact that F$^k$ = E$^k$ – TS$^k$ for each configuration and F = E – TS for the system, the total entropy of the system, S, can be defined as

$$S = \sum_{k=1}^{m} p^k S^k - k_B \sum_{k=1}^{m} p^k \ln p^k \quad \text{Eq. 5}$$

**Eq. 5** represents the central equation for Zentropy. Here, "Z" denotes the Zustandssumme, which translates to "sum over states" in German. Thus, this approach, which involves summing over the entropy of various states or configurations, is termed Zentropy [32].

Note that the equations in the previous zentropy publications of magnetic materials included the multiplicity of each configuration [23,25–28,30]. In this work, the multiplicities are not explicitly stated but are included in all the above summations.

The Helmholtz energy of each configuration, $F^k$, is given by

$$F^k = E^{k,0} + F^{k,vib} + F^{k,el} \qquad \text{Eq. 6}$$

where $E^{k,0}$ is the energy obtained from DFT calculations via energy-volume curves, $F^{k,vib}$ is the vibrational contribution to the free energy, which can be obtained from either phonon calculations via DFT or by using the Debye model, and $F^{k,el}$ is the thermal electronic contribution that can also be obtained via DFT. Here, we use phonon calculations first to obtain the phonon density of states (DOS) as a function of frequency, $g(v)$, which can then be used to calculate $F^{k,vib}$ using the following equation through numerical integration [35].

$$F^{k,vib} = k_B T \int_0^\infty \ln\left(2 \sinh \frac{hv}{2k_B T}\right) g(v) dv \qquad \text{Eq. 7}$$

where h is the Planck's constant. Here, we ignore the $F^{k,el}$ term as it is only important for metals [35] and negligible for $PbTiO_3$ as it is a wide band gap semiconductor. We can also evaluate the entropy, $S^k$, and the heat capacity at constant volume, $C_v^k$, of each configuration k

$$S^k = S^{k,vib} = k_B \int_0^\infty \left[\frac{hv}{2k_B T} \coth\left(\frac{hv}{2k_B T}\right) - \ln\left(2 \sinh \frac{hv}{2k_B T}\right)\right] g(v) dv \qquad \text{Eq. 8}$$

$$C_v^k = C_v^{k,vib} = k_B \int_0^\infty \left(\frac{hv}{2k_B T}\right)^2 \text{csch}^2\left(\frac{hv}{2k_B T}\right) g(v) dv \qquad \text{Eq. 9}$$

The Gibbs energy, G, and equilibrium volume, $V_c$, are obtained by the following equations [36].

$$G = \min_{V_c} [F + PV] \qquad \text{Eq. 10}$$

$$V_c = \arg\min_{V_c} [F + PV] \qquad \text{Eq. 11}$$

where P is the pressure, and V is the volume. Thus, **Eq. 10** and **Eq. 11** allow us to calculate constant pressure properties by using values that correspond to $V_c$, such as $p^k$ ($V_c$, T), and, in turn, evaluate other properties at these $p^k$ values, such as the configurational entropy, which is the second term in **Eq. 5**

$$S_{conf} = -k_B \sum_{k=1}^{m} p^k \ln p^k \qquad \text{Eq. 12}$$

We can also calculate the heat capacity of the system, $C_v$, which includes the configurational heat capacity, $C_{v,\,conf}$

$$C_v = \sum_{k=1}^{m} p^k C_v^k + C_{v,conf} \qquad \text{Eq. 13}$$

$$C_{v,conf} = \frac{1}{k_B T^2} \left[ \sum_{k=1}^{m} p^k (E^k)^2 - \left( \sum_{k=1}^{m} p^k E^k \right)^2 \right] \qquad \text{Eq. 14}$$

The present work uses the four-parameter Birch-Murnaghan (BM4) equation of state (EOS) [35] to fit the energy-volume curves from DFT, given by

$$E(V) = a + bV^{-\frac{2}{3}} + cV^{-\frac{4}{3}} + dV^{-2} \qquad \text{Eq. 15}$$

where a, b, c, and d are the fitting parameters. From these fitting parameters, one can determine the equilibrium volume, $V_0$, bulk modulus, $B_0$, and the derivative of the bulk modulus with respect to pressure, $B'_0$, as follows

$$V_0 = \sqrt{\frac{9bcd - 4c^3 - \sqrt{(c^2 - 3bd)(4c^2 - 3bd)^2}}{b^3}} \qquad \text{Eq. 16}$$

$$B_0 = \frac{2\left(27d + 14cV_0^{\frac{2}{3}} + 5bV_0^{\frac{4}{3}}\right)}{9V_0^3} \qquad \text{Eq. 17}$$

$$B_0' = \frac{243d + 98cV_0^{\frac{2}{3}} + 25bV_0^{\frac{4}{3}}}{81d + 42cV_0^{\frac{2}{3}} + 15bV_0^{\frac{4}{3}}} \qquad \text{Eq. 18}$$

## 3. Computational Methods

First-principles calculations were carried out using the VASP code [37,38] with the projector augmented-wave (PAW) method [39,40]. Valence electrons of $3s^2$ $3p^6$ $3d^2$ $4s^2$ for Ti, $5d^{10}$ $6s^2$ $6p^2$ for Pb, and $2s^2$ $2p^4$ for O were chosen. We used the more recently developed metaGGA $r^2$SCAN exchange-correlation functional [41], which is known to predict formation energies more accurately than PBEsol and SCAN and is computationally cheaper and more robust than SCAN [42]. We have also done some calculations using LDA [43–45] and PBEsol [46] to explore the sensitivity of the results with respect to different exchange-correlation functionals.

A comparison of ground state properties using different exchange-correlation functionals is shown in **Table 1**. With regards to the volume/formula unit (f.u.) or 5 atoms, the B1WC hybrid GGA [47] performs the best with less than 1% error underestimation compared to experimental results [48] since it was designed for ferroelectric materials being fitted to properties of the similar $BaTiO_3$. PBEsol performs the second best by overestimating the volume by 1.1%, SCAN overestimates the volume by 3.7%, LDA underestimates it by 3.7%, and $r^2$SCAN performs the worst by overestimating the volume by 4.8%. PBEsol has the best agreement with experiments with respect to the c/a ratio, underestimating it by less than 1%. This is followed by B1WC with an overestimation of 2.4% and LDA with an underestimation of 2.9%. The $r^2$SCAN and SCAN exchange-correlation functionals suffer from super tetragonality, overestimating the c/a ratio by 5% and 4.8%, respectively. With respect to the bulk modulus, LDA performs the best, with an underestimation of 9.2%. PBEsol and $r^2$SCAN underestimate the bulk modulus significantly, with an error of 59.7% and 65%, respectively. Lastly, the equilibrium energy difference between the cubic and tetragonal configurations of $r^2$SCAN most closely resembles the result of B1WC.

As seen above, different exchange-correlation functionals perform better for different properties. This issue of transferability of exchange-correlation functionals on predicting properties for a particular material, or even the same property across different materials, is a common problem in DFT [49]. For the purpose of this study, we chose r$^2$SCAN as the difference in equilibrium energy between the cubic and tetragonal configuration more closely resembles that of the more accurate hybrid B1WC compared to the other exchange-correlation functionals. As will be seen later, this energy difference correlates well with the domain wall energies, which is the most important parameter for predicting the transition temperature, $T_c$.

Similar to the work by Liu et al. [31], we consider three configurations: the ferroelectric ground state (FEG), 90° domain wall (90DW), and 180° domain wall (180DW), as illustrated in **Figure 2**. These domain wall structures are the only electrically neutral and mechanically compatible domain walls in the tetragonal phase [50]. They have also been observed experimentally [51–53]. For the FEG configuration, the directions of the polarizations, as indicated by the shortest Ti-O bonds, are all aligned in the same direction, as shown in **Figure 2a**. **Figure 2b** shows a head-to-tail 90DW structure where the polarizations on either side of the domain wall on the {101} plane are almost perpendicular to one another. This head-to-tail 90DW is stable as opposed to the unstable head-to-head or tail-to-tail 90DW [54]. **Figure 2c** shows a Pb-centered 180DW structure with the domain wall on the {100} plane, where the polarizations on either side of the domain wall are parallel and opposite. The head-to-head and tail-to-tail 180DW structures [55,56] have been found to be highly charged with high domain wall energies and are thus not considered in this work due to likely having almost negligible $p^k$ values. Likewise, this work does not consider high-energy Ti-centered 180DW structures [57]. The same logic applies to the standard cubic phase, where its energy-volume curve can be shown in **Figure S1** of the Supplemental Material [58]. Moreover, as shown in **Figure S2**, the significant number of imaginary phonons does not allow us to compute the free energy of that configuration.

In applying zentropy, the multiplicity of each configuration needs to be considered in the sums of all the equations above. For the FEG configuration (**Figure 3a**), the first domain has six energetically equivalent orientations where the polarization and the corresponding c-axis direction [001] can be aligned along the ±x, ±y, or ±z axes. The second domain can be placed on one of the other four sides of the remaining two axes and must have its polarization pointing in the same direction as the first. Thus, the total number of equivalent configurations is 6 × 4 × 1 = 24. Likewise, the 180DW configuration has the same number of equivalent configurations but with the condition that the polarization of the second domain is in the opposite direction of the first (**Figure 3b**), leading to the same total of 6 × 4 × 1 = 24. As previously mentioned by Liu et al. [31], for the 90DW configuration, the second domain has four degrees of freedom for the polarization direction. However, only one of these directions leads to the head-to-tail configuration (**Figure 3c**). One other direction leads to the unstable head-to-head configuration (**Figure 3d**). In contrast, the other two directions lead to either the head-to-in configuration (**Figure 3e**) or the head-to-out configuration (**Figure 3f**) not yet reported in the literature, which are not considered in the present work. Consequently, the multiplicity of 90DW becomes 6 × 4 × 1 = 24 by only including the stable head-to-tail 90DW rather than 6 × 4 × 4 = 96 from Liu et al. [31]. Thus, the multiplicities used for FEG, 90DW, and 180DW are 24:24:24 or a reduced ratio of 1:1:1.

For the energy-volume curves, 10 × 1 × 1 or 50-atom supercells of the perovskite unit cell were used for all the configurations in accordance with the convergence of the domain wall energies presented in the literature [57,59,60]. The convergence criterion for the electronic self-consistency loop (EDIFF) and the ionic relaxation loop (EDIFFG) was set to $1 \times 10^{-6}$ eV and 0.01 eV/Å, respectively. The cutoff energy (ENCUT) was taken as 500 eV, and a 1 × 6 × 6 Γ-centered k-point mesh was used. Two consecutive relaxations (IBRION = 2) were performed at a fixed volume (ISIF = 4), allowing the ionic positions and cell shape to relax (which allows the c/a ratio to change) with the Gaussian-smearing method (ISMEAR = 0; SIGMA = 0.05), followed by a static calculation (IBRION = -1; NSW = 0) with the tetrahedron method with Blöchl corrections (ISMEAR = -5). Non-spherical contributions from the gradient corrections inside the PAW spheres were also included (LASPH = True).

For the phonon calculations, 10 × 2 × 2 or 200-atom supercells of the perovskite unit cell replicated from the relaxed configurations of the energy-volume calculations were used for all configurations. This supercell size is to ensure that we obtain reasonable phonon results by limiting atomic interactions between displaced atoms for the finite-displacement supercell approach, as the dimensions of the 50-atom supercell used for the energy-volume curves are too small in the y and z-directions [36]. The second-order derivatives of the total energy with respect to the position of the ions were computed in VASP with symmetry applied to reduce the number of displacements (IBRION = 6). We've determined that the convergence criterion for the electronic self-consistency loop (EDIFF) of $1 \times 10^{-4}$ eV was sufficient with the Gaussian-smearing method applied. For this larger supercell, a 1 × 3 × 3 Γ-centered k-point mesh was used. The remaining VASP settings were the same as the ones used for the energy-volume curves.

The phonon DOS was obtained by using YPHON [61], which uses a mixed-space approach that calculates the force constants in real space and the dipole-dipole interactions in reciprocal space. Example VASP input files used to generate the results for this paper can be found in the Supplemental Material [58]. The results from the energy-volume and phonon calculations were then used as inputs to zentropy.

## 4. Results and Discussion

Figure 4 shows the energy-volume curves from DFT calculations for 50-atom supercells of the FEG, 90DW, and 180DW configurations using r$^2$SCAN. These curves represent the $E^{k,0}$ term in **Eq. 6**. Seven to eight DFT data points were used to fit the BM4 EOS using **Eq. 15 – Eq. 18** with volumes ranging from 610 to 690 Å$^3$. For volumes smaller than 610 Å$^3$, the 180DW configuration was found to be unstable and relaxed to the FEG configuration.

**Table 1** shows the equilibrium volumes, equilibrium energies, bulk modulus, and derivative of the bulk modulus with respect to pressure, B', obtained from the EOS fitting. The equilibrium volume for FEG (656.52 Å$^3$) is larger than that of 90DW (644.97 Å$^3$), which is larger than that of 180DW (633.33 Å$^3$). On the other hand, the equilibrium energy for FEG (-999.38 eV) is lower than that of 90DW (-999.15 eV), which is lower than 180DW (-998.91 eV). These properties, when put into the zentropy approach, will show that this material has a negative thermal expansion, as observed in experiments [6]. The bulk modulus increases in the order of FEG, 90DW, and 180DW, with values of 36.39, 38.63, and 64.33 GPa, respectively, but is much lower compared to the experimental value of 104 GPa for the tetragonal phase at room temperature [62,63]. The reason for this large deviation from experiments is unclear. It requires future investigation as r$^2$SCAN has been found to show good agreement for the bulk moduli of various transition metals [64]. We also note that the high B' values of 7.39 – 27.55 are unique for PbTiO$_3$, significantly different from the commonly observed B' values of 4 – 6 for our previous work on metals [35]. The energy-volume curves using LDA and PBEsol and their associated EOS parameters are shown in **Figure S3**, **Table S1**, and **Table S2** in the Supplemental Material [58].

At their respective equilibrium volumes, the equilibrium energy difference between 90DW and FEG is 4.53 meV/atom, while the difference between 180DW and FEG is 9.35 meV/atom. However, the domain wall energies are usually calculated using the energies at the fixed equilibrium volume of FEG [53,54]. **Figure 5** shows the energy difference as a function of volume for 90DW and 180DW using FEG as the reference state. Fixing the volume at the equilibrium volume of FEG (658.19 Å$^3$) using r$^2$SCAN, the domain wall energies for 90DW and 180DW are 91.1 and 281.5 mJ/m$^2$, respectively. We have also conducted some calculations using LDA and PBEsol for comparison. The domain wall energies for 90DW and 180DW are 57.6 and 193.5 mJ/m$^2$ using PBEsol and 27.2 and 117.3 mJ/m$^2$ using LDA, respectively. The energy differences used to calculate the domain wall energies were taken at their respective fixed equilibrium volumes for FEG (632.92 Å$^3$ for PBEsol and 603.35 Å$^3$ for LDA).

The LDA domain wall energy values are in good agreement with 29.2 mJ/m$^2$ for 90DW from Shimada et al. [59] and 124 mJ/m$^2$ for 180DW from Behera et al. [57] for the same supercell size. Our domain wall energies are lower as we allow both the atomic positions and cell shape to relax, while these previous studies [57,59] only allowed the atomic positions to relax. The larger domain wall energies of 35 and 132 mJ/m$^2$ for the 90DW and 180DW, respectively, reported by Meyer and Vanderbilt [60] are because they used the ultrasoft-pseudopotential approach with LDA rather than the PAW method used here and by Shimada et al. [59] and Behera et al. [57]. The large difference in the domain wall energies comparing r$^2$SCAN, PBEsol, and LDA highlights the sensitivity of material properties to the choice of the exchange-correlation functional. We have chosen to use the metaGGA r$^2$SCAN instead of LDA or PBEsol as it is a higher level of theory, offering a higher accuracy than GGA and LDA across most of the chemical space while being computationally tractable [42]. As will be shown later, r$^2$SCAN also leads to a better prediction of T$_c$.

**Figure 6** shows the phonon DOS using r$^2$SCAN for FEG (**Figure 6a**), 90DW (**Figure 6b**), and 180DW (**Figure 6c**). The phonon DOS using 200-atom supercells were normalized to 50-atom supercells in **Figure 6**, where the area under the curve is equal to 3N, with N being the number of atoms in the supercell as processed by YPHON [61]. The phonon DOS were obtained at four fixed volumes for FEG (630, 640, 650, and 660 Å$^3$), 90DW (630, 639, 651.8, and 664.6 Å$^3$), and 180DW (630, 640, 650, and 660 Å$^3$). The phonon calculations were run at these fewer fixed volumes compared to the energy-volume curves due to the higher computational costs of these calculations, especially for the 90DW and 180DW configurations with much lower symmetry than FEG, therefore requiring significantly more displacements in the finite displacement supercell approach.

**Eq. 7** was then used to calculate $F^{k,vib}$ at these volumes as a function of temperature from 10 to 1000 K. As seen in **Figure 6**, the phonon DOS for all configurations for frequencies less than 3 THz are almost identical, which leads to $F^{k,vib}$ being approximately constant as a function of volume for each temperature. To illustrate this, we plotted $F^{k,vib}$ for three fixed temperatures of 600 K, 700 K, and 800 K, as shown in **Figure 7a**. This result justifies using fewer volumes for the phonon calculations compared to the energy-volume curves. For each configuration at each temperature, the calculated $F^{k,vib}$ (solid circles) was fitted to a linear equation (solid lines) to obtain the remaining values as a function of volume. We found that for all fixed temperatures, $F^{k,vib}$ for FEG is always the least negative, followed by 90DW and 180DW.

With $E^{k,0}$ and $F^{k,vib}$ obtained for each configuration, the free energy of each configuration, $F^k$, was predicted by summing these two terms as seen in **Eq. 6**. The Gibbs energy of each configuration as a function of temperature at P = 0 GPa could be calculated using **Eq. 10**. The difference in Gibbs energy with FEG as the reference state is plotted in **Figure 7b**, showing that 90DW becomes more stable than FEG from approximately 765 – 990 K, after which the 180DW becomes more stable, due to its lower $F^{k,vib}$ causing its Gibbs energy to decrease more rapidly as a function of temperature compared to 90DW.

The partition function of each configuration, $Z^k$, and the partition function of the system, $Z$, were then obtained via **Eq. 1** and **Eq. 2**. Next, the probability of each configuration, $p^k$, was obtained via **Eq. 3**, which was used to obtain the total Helmholtz energy, F, of the system using **Eq. 4** and subsequently the system equilibrium volumes, $V_c$, at P = 0 using **Eq. 11**. **Figure 7c** shows $p^k$ which corresponds to these $V_c$ at P = 0 GPa as a function of temperature. Similar to our previous zentropy work on magnetic materials [23,25–30], we define the Curie temperature, $T_c$, to occur when $p^k = 0.5$ for FEG and the sum of all other configurations (90DW + 180DW). Using this criterion, the predicted $T_c$ is 716 K using r$^2$SCAN, which is 47 K below the experimental $T_c$ of 763 K [6]. We have also run phonon calculations and applied our zentropy approach using LDA and PBEsol, which predicts $T_c$ = 222 K and $T_c$ = 436 K, respectively, severely underestimating the experimental $T_c$ and further justifying our choice for using r$^2$SCAN as the exchange-correlation functional. The same plots for LDA and PBEsol are shown in **Figure S4** and **Figure S5**, respectively, in the Supplemental Material [58].

The value of $p^k = 0.5$ is one of our criteria for determining the phase transition, which is similar to the percolation threshold for a simple square system [65]. Other criteria have been used in our previous publications, including the peak value of the configurational heat capacity, $C_{v,\ conf}$ [25] computed using **Eq. 14**, and plotted in **Figure 8a**, which shows a peak value at 719 K corresponding well to 716 K from the $p^k = 0.5$ criterion. In **Figure 8b**, the configurational entropy, $S_{conf}$, computed using **Eq. 12**, is plotted as a function of temperature with a peak value at 763 K, which is identical to the experimental observations, indicating that this may be a better criterion for phase transition and is worth of further exploration in other systems.

Lastly, **Figure 9** shows the normalized equilibrium volume as a function of temperature in the present work with the experimental values of Shirane and Hoshino [6] superimposed. There are two apparent similarities. One is that the change in volume over the temperature range is minimal, less than 4% for this work and less than 2% for experiments. The second similarity is that the most significant drop in volume occurs near $T_c$. However, the experimental results show a sharper drop in volume compared to the present work. A similar difference was observed in a previous zentropy study involving $Fe_3Pt$ [28], which can be attributed to the small supercell size used in DFT calculations. The experiment also shows negative thermal expansion from 303 K to $T_c$, after which the material undergoes a slight positive thermal expansion. In contrast, the present work shows a slight positive thermal expansion from 0 K to 495 K, after which it only shows negative thermal expansion and no positive thermal expansion after $T_c$. The equilibrium volumes of each configuration and the system using LDA, PBEsol, and $r^2$SCAN are shown in **Figure S6** of the Supplemental Material [58].

These differences highlight some limitations of our current approach to ferroelectric materials. In this work, we only computed the free energy as a function of volume for each configuration, as shown in **Figure 4**, where the c/a ratio relaxes to a different value at each volume. This result is shown in **Table 3**, which reports different average c, a, and c/a ratio values in the x-z plane for each configuration at various fixed volumes at 0 K. Here, we present the average c/a ratio as the c and a bond lengths vary in the 90DW and 180DW supercells. Applying this to our zentropy approach, we can obtain the equilibrium volume of the system as a function of temperature, as shown in **Figure 9**, but unable to resolve the c/a ratio of the system at each temperature. To do so, one would have to compute free energy not just as a function of volume but also as a function of c and a, which is a subject of future work.

Furthermore, only three configurations considered in the present work with 50-atom or 10x1x1 supercells were selected from the literature and were constructed for the convergency of domain wall energy [57,59]. On the other hand, the AIMD simulations conducted by Fang et al. [10] showed more complex configurations in terms of local polarization configurations. For the similar material $BaTiO_3$, the cubic phase with intrinsic local symmetry breaking and local Ti distortion was found to be more stable compared to the centrosymmetric cubic phase at 0 K [66,67], which appear similar to the results of Fang et al. [10]. Thus, it is necessary to generate and include many types of these symmetry-breaking configurations in $PbTiO_3$ in order to fully reproduce all experimentally observed features in addition to the transition temperature.

However, the number of possible configurations for a given supercell is vast. For example, for a 40-atom or 2x2x2 supercell with 8 Ti atoms with a cubic shape, the total number of possible configurations is $6^8 = 1,679,616$, where 6 represents the number of polarization directions along the ± x, y, and z directions. It is the subject of future work to find the unique number of configurations out of this possible total and then perform high-throughput DFT calculations to compute the free energy as a function of c and a using our soon-to-be revised Density Functional Theory Toolkit (DFTTK) [68] for high-throughput DFT calculations and transfer learning of neural network machine learning models [69].

## 5. Summary

In summary, the present work comprises our new zentropy study with the ferroelectric material $PbTiO_3$ using the free energies of the FEG, 90DW, and 180DW configurations predicted by DFT. Using the new metaGGA $r^2$SCAN exchange-correlation functional, zentropy predicts $T_c$ to be 716 K according to the $p^k = 0.5$ criterion, in good agreement with the experimental value of 763 K. This value also agrees with the heat capacity criterion, where the peak of the configurational heat capacity, $C_{v,\,conf}$, occurs at 719 K, and the peak of the configurational entropy, $S_{conf}$ at 763 K. The equilibrium volume as a function of temperature agrees well with experiments regarding the minimal change in volume (< 4%) over the temperature range of 200 – 900 K and that the most significant drop in volume occurs near $T_c$. However, due to the limitations of the number and types of configurations adopted from the literature, the temperature ranges for negative and positive thermal expansion could not be fully captured. It is the subject of future work to define a new set of configurations and compute the free energy as a function of a and c to accurately predict the cubic phase for $T > T_c$, the temperature range of negative thermal expansion, and the temperature-pressure phase diagram.

# 6. Acknowledgments

This work is funded by the Department of Energy (DOE) through Grant No. DE-SC0023185. First-principles calculations were conducted at the Roar Collab cluster at the Pennsylvania State University, the Bridges-2 supercomputer at the Pittsburgh Supercomputing Center, and the Setonix supercomputer at the Pawsey Supercomputing Research Centre. The authors would like to thank Dr. Dino Spagnoli from the University of Western Australia for allowing us to use computing time for this project on Setonix.

**Figures**

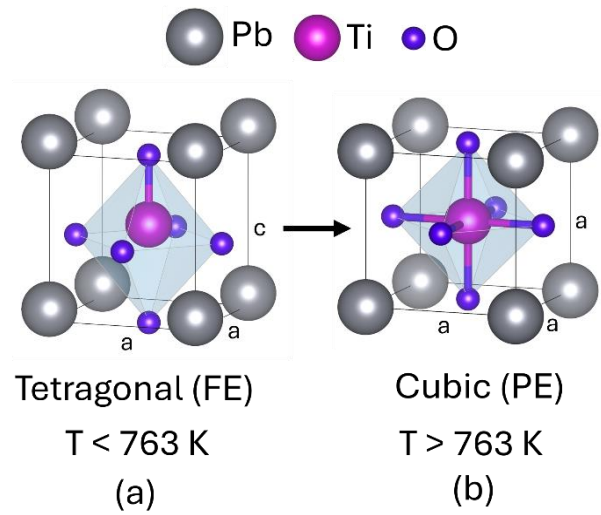

**Figure 1.** The unit cell of PbTiO$_3$ according to XRD measurements. (a) A tetragonal ferroelectric (FE) phase below 763 K. (b) Transition to a cubic paraelectric (PE) phase at 763 K.

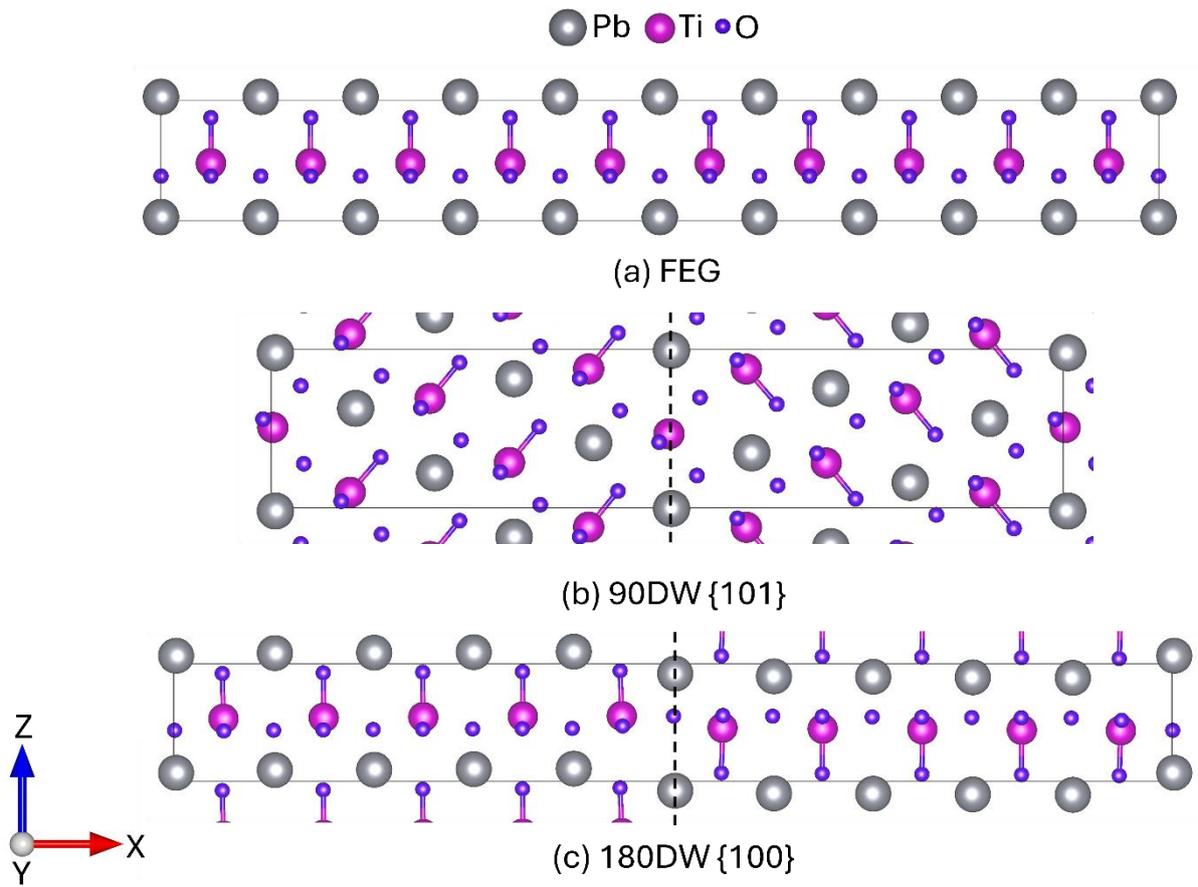

**Figure 2.** 50-atom supercells of the (a) ferroelectric ground state (FEG), (b) 90° domain wall (90DW), and (c) 180° domain wall (180DW). Only the shortest Ti-O bonds are shown to illustrate the direction of the polarization.

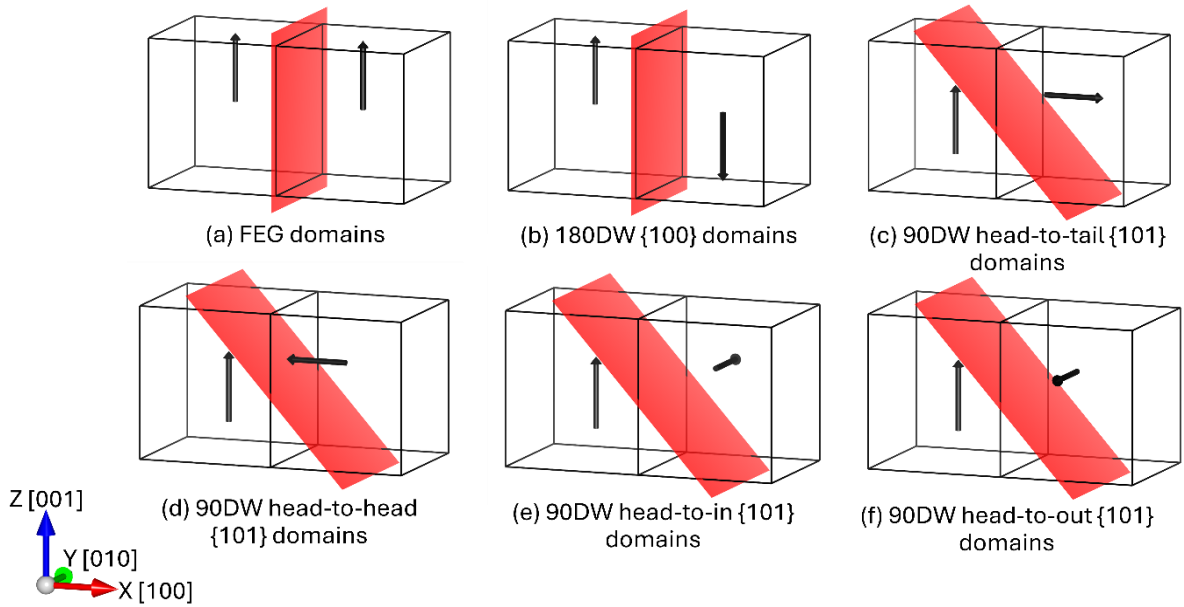

**Figure 3.** The domains used to calculate the multiplicities of FEG, 90DW, and 180DW. The domains used in this work are the (a) FEG domains, (b) 180DW {100} domains, and the (c) 90DW head-to-tail {101} domains. The domains not considered in this work are the (d) 90DW head-to-head {101} domains, 90DW head-to-in {101} domains, and the 90DW head-to-out {101} domains.

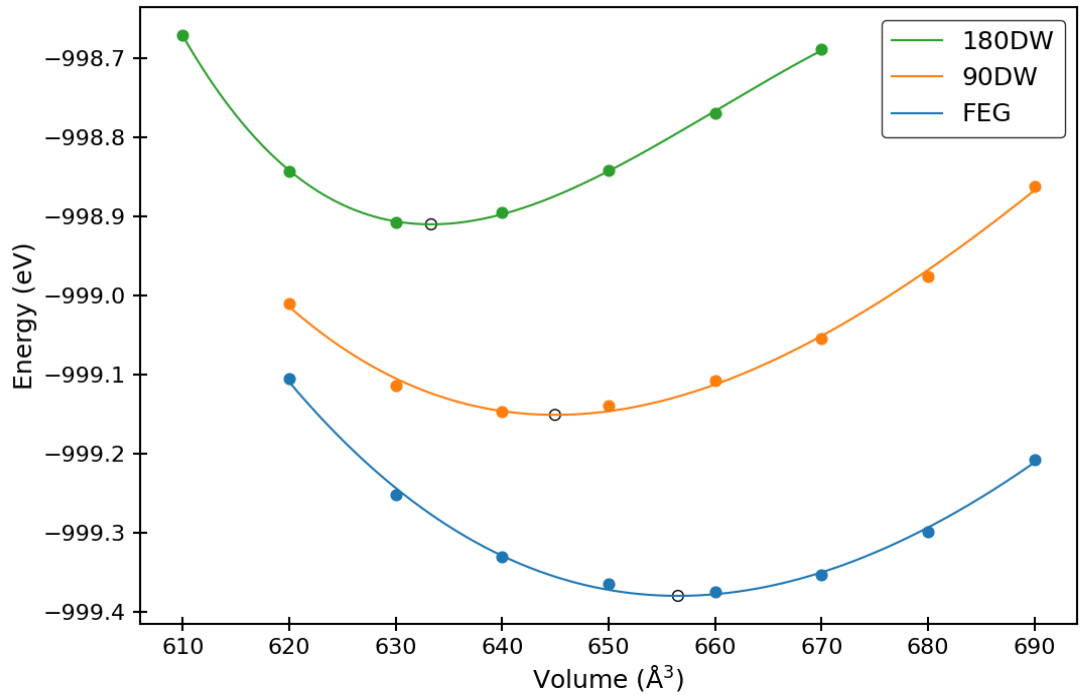

**Figure 4.** Energy-volume curves of FEG, 90DW, and 180DW. The solid circles are the results from DFT calculations for 50-atom supercells using the r$^2$SCAN exchange-correlation functional. These were used to fit the four-parameter Birch-Murnaghan (BM4) equation of state (EOS) to obtain the solid curves. The black open circles represent the minimum energy and volumes from the fitting.

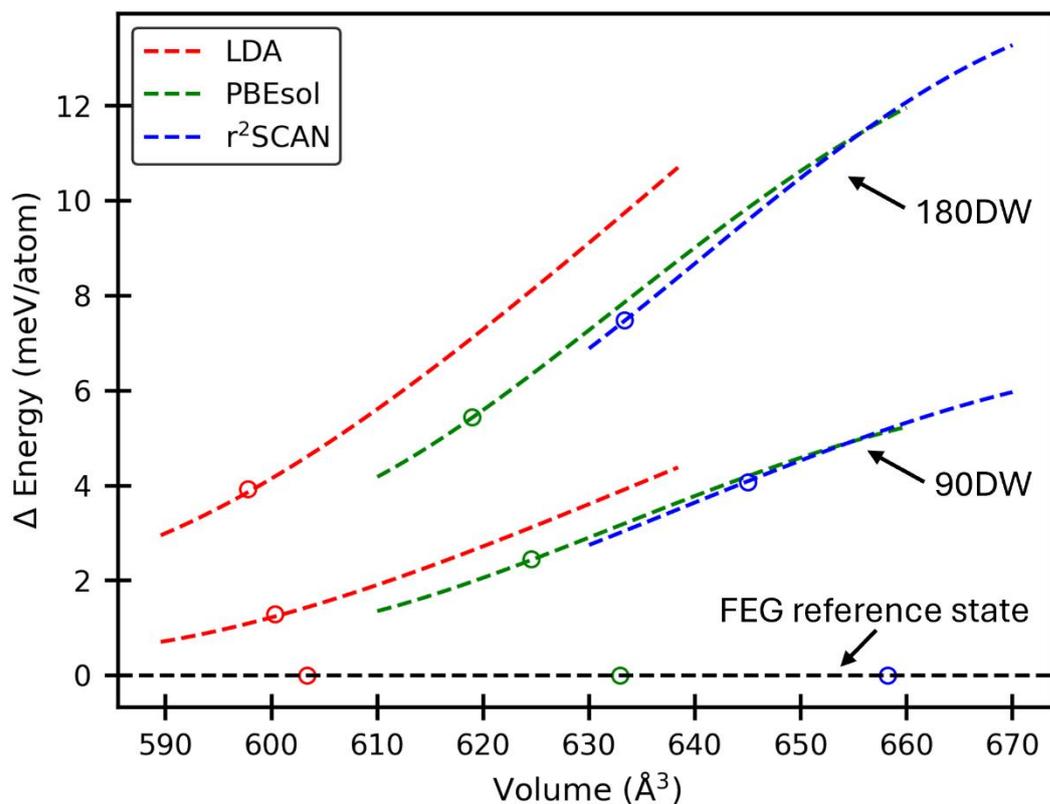

**Figure 5.** Energy difference as a function of volume for 90DW and 180DW with FEG as the reference state using the LDA, PBEsol, and r²SCAN exchange-correlation functionals. The open circles represent the relative equilibrium energies and volumes determined from fitting DFT results to the four-parameter Birch-Murnaghan (BM4) equation of state.

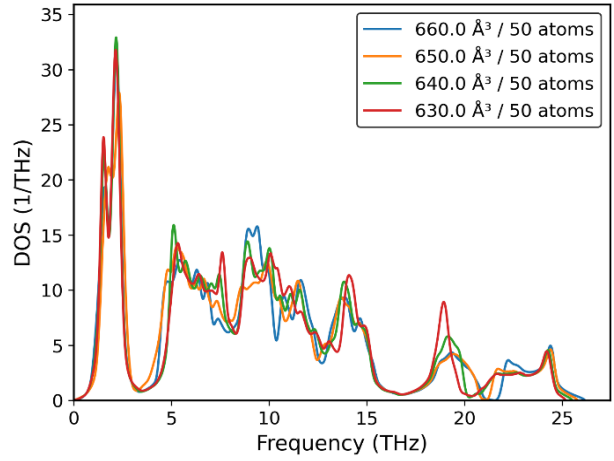

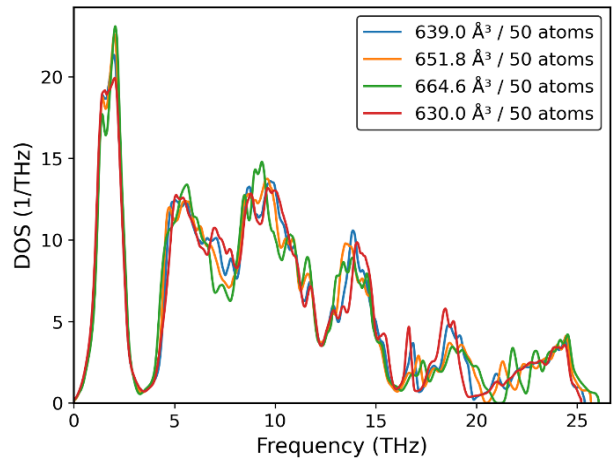

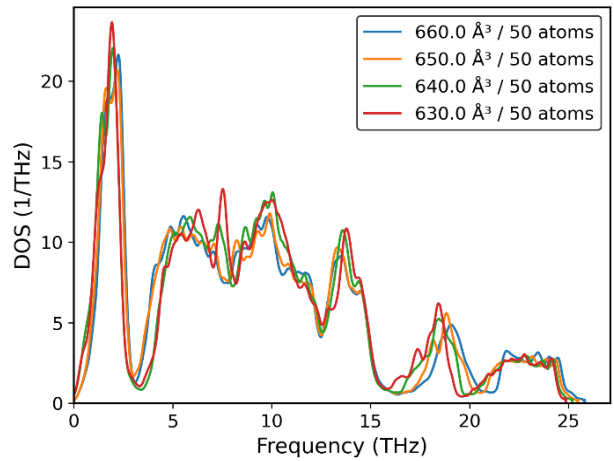

**Figure 6.** Phonon density of states (DOS) obtained from DFT using r$^2$SCAN for the (a) FEG, (b) 90DW, and (c) 180DW configurations at multiple fixed volumes using 200-atom supercells, normalized to 50-atom supercells where the area under the curve is equal to 3N, and N is the number of atoms.

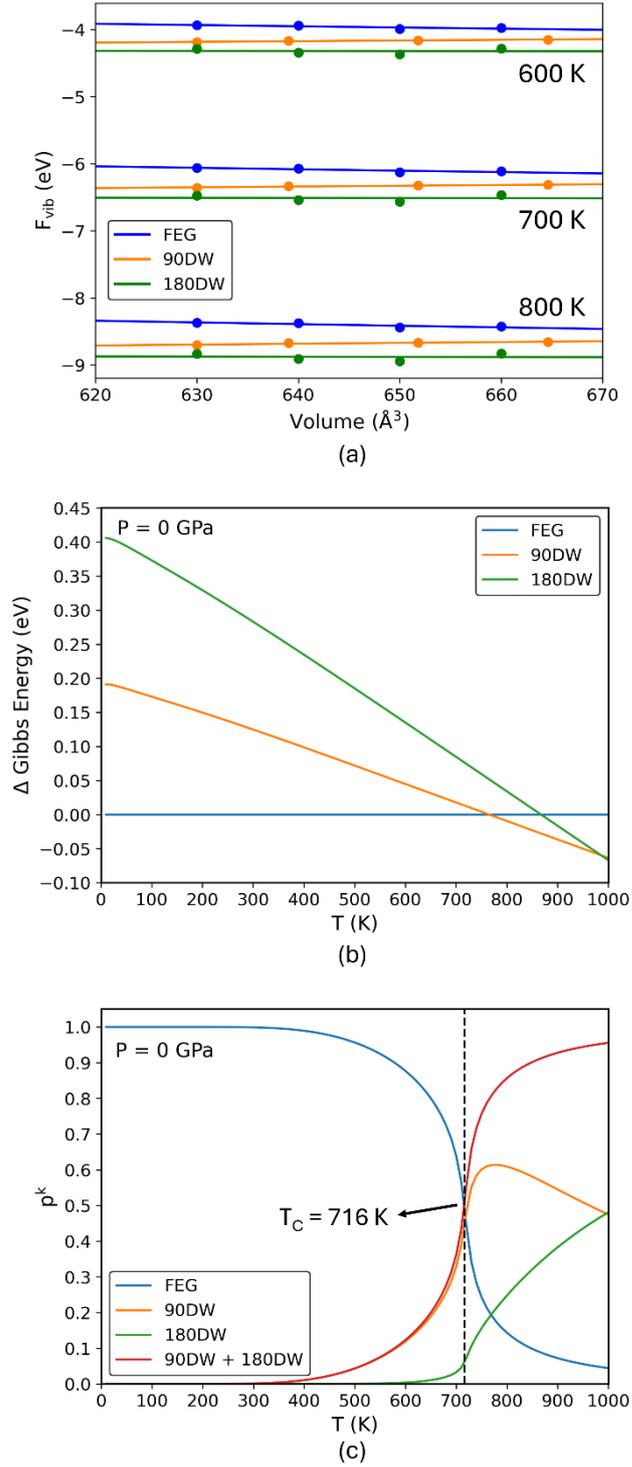

**Figure 7.** (a) The vibrational contribution to the free energy for 50-atom supercells, $F^{k,vib}$, calculated from the phonon DOS using $r^2$SCAN for three fixed temperatures: 600 K, 700 K, and 800 K. (b) The difference in the Gibbs energy for 50-atom supercells at P = 0 GPa with FEG as the reference state. (c) The probabilities, $p^k$, as a function of temperature at P = 0 GPa with the Curie Temperature, $T_c$, defined by $p^{FEG} = 0.5$.

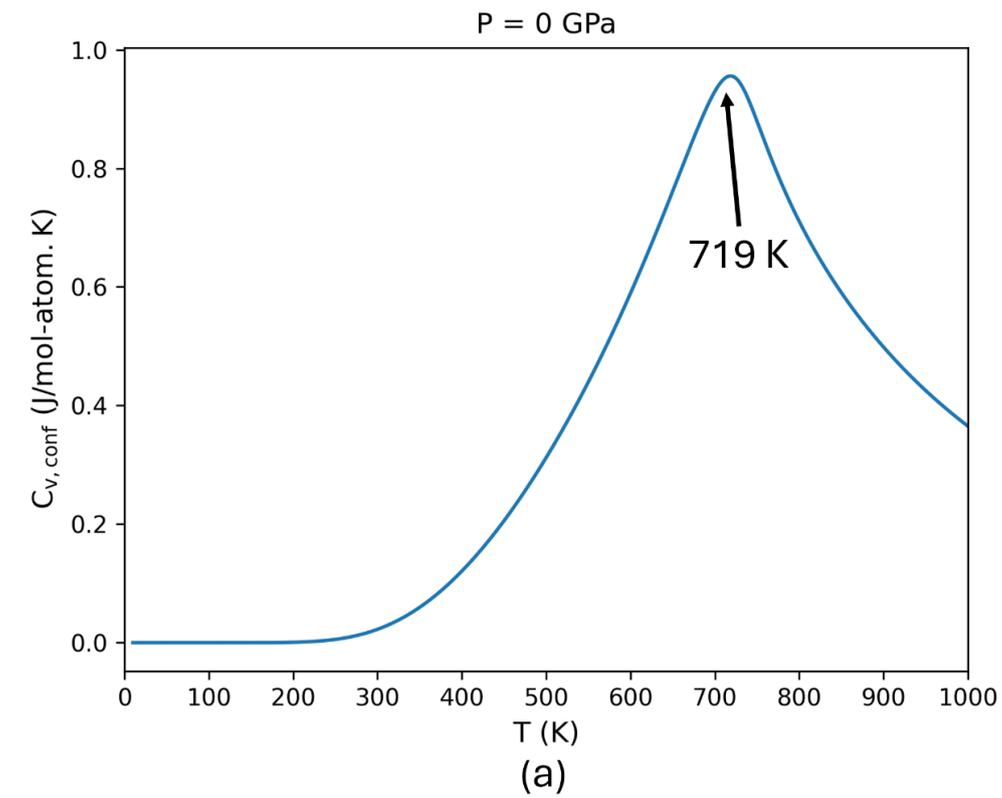

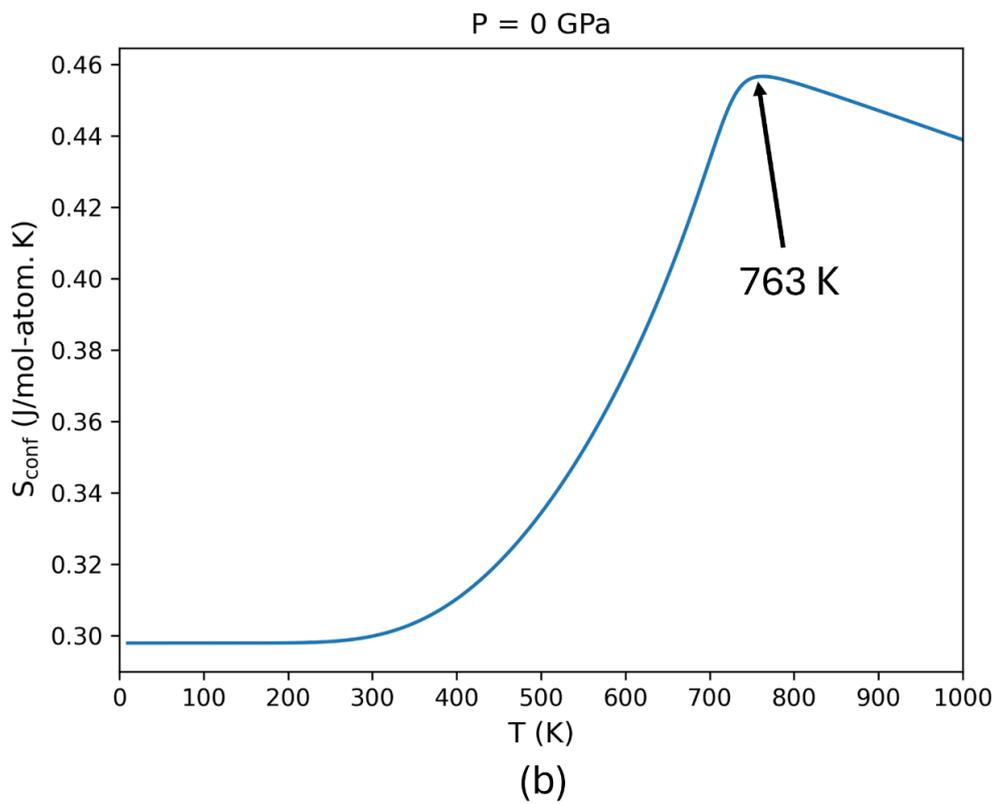

**Figure 8.** (a) Configurational heat capacity, $C_{v,\ conf}$, with a peak value at 719 K, and (b) Configurational entropy, $S_{conf}$, with a peak value at 763 K as a function of temperature.

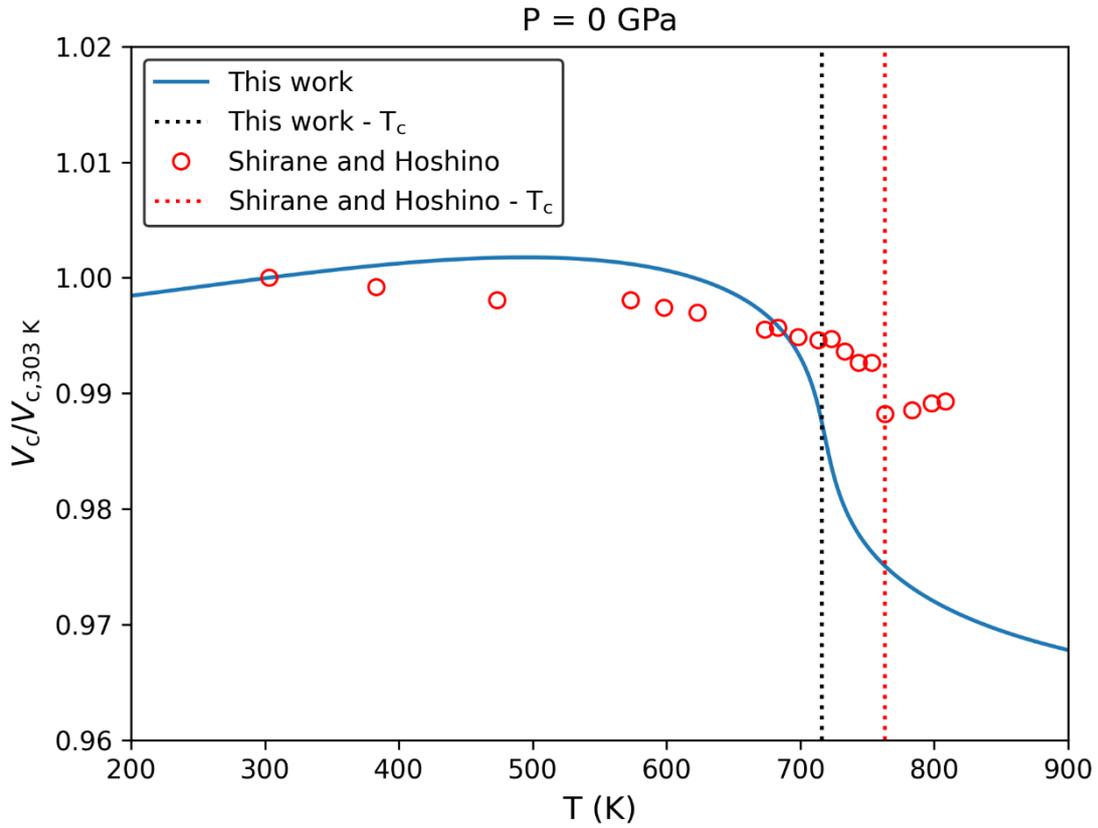

**Figure 9.** Relative volume increase, $V_c/V_{c,303\,K}$, as a function of temperature at P = 0 GPa. The $V_{c,303\,K}$ from this work is 66.03 Å³/5 atoms compared to the $V_{c,303\,K}$ of 62.77 Å³/5 atoms from Shirane and Hoshino [6]. The blue solid line represents the result from this work, while the open red circles represent experimental data from Shirane and Hoshino [6]. The black and red dotted lines represent $T_c$ predicted from this work and experiments, respectively.

**Table 1.** Comparison of the ground state FEG properties – equilibrium volume/formula unit (f.u.) or 5 atoms (Å³), c/a ratio, bulk modulus, B (GPa), and the equilibrium energy difference/f.u. between the FEG and standard cubic phase, ΔE (meV) - using the LDA, PBEsol, r$^2$SCAN, SCAN, and B1WC exchange-correlation functionals.

| Property | LDA | PBEsol | r$^2$SCAN | SCAN | B1WC | Expt. |
|---|---|---|---|---|---|---|
| **Volume/f.u. (Å³)** | 60.3352 | 63.2924 | 65.6515 | 64.9 [48] | 62.4 [48] | 62.6 [48] |
| **c/a ratio** | 1.04 | 1.066 | 1.125 | 1.122 [48] | 1.097 [48] | 1.071 [48] |
| **B (GPa)** | 94.39 | 41.87 | 36.39 | - | - | 104 [62,63] |
| **ΔE/f.u. (meV)** | 48 | 75 | 109.1 | 122.7 [48] | 110.6 [48] | - |

**Table 2.** Equilibrium volumes (Å³), equilibrium energies (eV), bulk modulus, B (GPa), and derivative of the bulk modulus with respect to pressure, B', obtained from fitting r²SCAN DFT data points for 50-atom supercells of FEG, 90DW, and 180DW to the four-parameter Birch Murnaghan (BM4) equation of state (EOS), at 0 K.

| | r²SCAN | | | |
|---|---|---|---|---|
| Configuration | Volume (Å³/50 atoms) | Energy (eV/50 atoms) | B (GPa) | B' |
| FEG | 656.5153 | -999.3799 | 36.39 | 7.39 |
| 90DW | 644.9687 | -999.1512 | 38.63 | 11.36 |
| 180DW | 633.3287 | -998.9104 | 64.33 | 27.55 |

**Table 3.** The average c, a, and c/a ratio in the x-z plane for each configuration at various relaxed, fixed volumes per 50-atom supercells using r$^2$SCAN.

| | r$^2$SCAN | | | |
|---|---|---|---|---|
| Configuration | Volume (Å$^3$/50 atoms) | Avg. c (Å) | Avg. a (Å) | Avg. c/a ratio |
| FEG | 690 | 4.686 | 3.837 | 1.221 |
| | 680 | 4.606 | 3.842 | 1.199 |
| | 670 | 4.528 | 3.847 | 1.177 |
| | 660 | 4.438 | 3.856 | 1.151 |
| | 650 | 4.344 | 3.868 | 1.123 |
| | 640 | 4.248 | 3.882 | 1.094 |
| | 630 | 4.168 | 3.888 | 1.072 |
| | 620 | 4.096 | 3.891 | 1.053 |
| 90DW | 690 | 4.743 | 3.801 | 1.248 |
| | 680 | 4.669 | 3.802 | 1.228 |
| | 670 | 4.579 | 3.810 | 1.202 |
| | 660 | 4.478 | 3.823 | 1.171 |
| | 650 | 4.370 | 3.842 | 1.137 |
| | 640 | 4.263 | 3.864 | 1.103 |
| | 630 | 4.170 | 3.879 | 1.075 |
| | 620 | 4.088 | 3.892 | 1.051 |
| 180DW | 670 | 4.484 | 3.892 | 1.152 |
| | 660 | 4.380 | 3.903 | 1.122 |
| | 650 | 4.286 | 3.911 | 1.096 |
| | 640 | 4.197 | 3.917 | 1.071 |
| | 630 | 4.119 | 3.921 | 1.050 |
| | 620 | 4.057 | 3.917 | 1.036 |
| | 610 | 4.007 | 3.907 | 1.025 |